\documentclass[aps,floatfix,superscriptaddress,reprint,10pt,prl]{revtex4-1}
\usepackage{bbm}
\usepackage{bm}
\usepackage{amsmath}
\usepackage{amssymb}
\usepackage{empheq}
\usepackage{graphicx}
\usepackage{mathrsfs}
\usepackage{amsfonts}
\usepackage{amsthm}
\usepackage{color}
\usepackage{bigints}
\usepackage{txfonts}
\usepackage{hyperref}
\usepackage{color}
\usepackage{appendix}
\usepackage{multirow}
\usepackage{makecell}

\hypersetup{
	unicode=false,               pdftoolbar=true,             pdfmenubar=true,             pdffitwindow=false,          pdfstartview={FitH},         pdftitle={My title},         pdfauthor={Author},          pdfsubject={Subject},        pdfcreator={Creator},        pdfproducer={Producer},      pdfkeywords={keyword1} {key2} {key3},      pdfnewwindow=true,           colorlinks=false,            linkcolor=red,               citecolor=green,             filecolor=magenta,           urlcolor=cyan           }
\makeatletter \tolerance = 10000 \tolerance = 10000
\makeatother

\begin{document}
	\title{Unconventional SOC-activated Altermagnetic Spin Splitting and Half-Angle Locking of Spin Transport}
	
	\author{Weiwei Chen}
	\affiliation{Department of Physics, Hefei University of Technology, Hefei, Anhui 230601, China}
	
	\author{Ziyu Zhou}
	\affiliation{Department of Physics, Hefei University of Technology, Hefei, Anhui 230601, China}
	
	\author{Pan Gao}
	\affiliation{Department of Physics, Hefei University of Technology, Hefei, Anhui 230601, China}
	
	\author{Yanan Pan}
	\affiliation{Department of Physics, Hefei University of Technology, Hefei, Anhui 230601, China}
	
	\author{Zhongjun Li}
	\affiliation{Department of Physics, Hefei University of Technology, Hefei, Anhui 230601, China}
	
	\author{Ye Yang}
	\thanks{Contact author. E-mail: yangye@hfut.edu.cn}
	\affiliation{Department of Physics, Hefei University of Technology, Hefei, Anhui 230601, China}
	
	\author{Jianhui Zhou}
	\thanks{Contact author. E-mail: jhzhou@hmfl.ac.cn}
	\affiliation{Anhui Provincial Key Laboratory of Low-Energy Quantum Materials and Devices, High Magnetic Field Laboratory, HFIPS, Chinese Academy of Sciences, Hefei, Anhui 230031, People’s Republic of China}
	
	\author{Rui Wang}
	\affiliation{Department of Physics and Chongqing Key Laboratory for Strongly Coupled Physics, Chongqing University, Chongqing 400044, China}
	\affiliation{Center of Quantum Materials and Devices, Chongqing University, Chongqing 400044, China}
	
	\begin{abstract}
		Conventional altermagnetic spin splitting arises from staggered exchange and sublattice-dependent anisotropic hopping without requiring spin–orbit coupling (SOC). 
		Here we derive a low-energy effective model showing that SOC and staggered exchange can jointly activate additional spin splitting on momentum planes where the nonrelativistic contribution vanishes.
		In hexagonal $\alpha$-MnTe, this SOC-activated contribution produces d-wave spin splitting near $\Gamma$ with out-of-plane spin polarization.
		Its leading angular pattern counter-rotates by half the in-plane N\'eel-vector rotation angle.
		First-principles spin-transport calculations reveal a corresponding approximate half-angle locking between the N\'eel-vector orientation and the electric-field directions maximizing the signed longitudinal and transverse spin conductivities.
		Analytical transport calculations based on the effective model trace this locking to the same angular coupling between momentum and the N\'eel vector that governs the spin-splitting pattern.
		Our findings connect the microscopic mechanism of SOC-activated altermagnetic spin splitting to directional spin transport, providing a theoretical basis for probing N\'eel-vector orientation and controlling spin-current generation.
	\end{abstract}
	

	\maketitle

\textit{Introduction.---} Altermagnets are a new class of compensated collinear magnets that exhibit momentum-dependent spin splitting without a net magnetic moment, offering a promising platform for spin-polarized transport and low-dissipation spintronics~\cite{Smejkal2022Beyond,Smejkal2022prxEmerging,Mazin2022prx,Hayami2019jpsj}. 
In the conventional nonrelativistic picture, this splitting arises from the interplay between staggered exchange and sublattice-dependent anisotropic hopping permitted by reduced local point symmetry~\cite{Hayami2019jpsj,Roig2024prb}. Spin–orbit coupling (SOC) is not required for this mechanism, and crystal symmetry constrains the resulting spin-splitting patterns to forms such as d, g, or i wave~\cite{Pan2026arxiv,Smejkal2022Beyond,Hayami2019jpsj}. Altermagnetic spin splitting and spin textures have been experimentally probed by photoemission spectroscopy~\cite{Lee2024PRL,Krempasky2024nature} and helical-tunneling microscopy~\cite{wang2026nature}.

The absence of SOC from this mechanism, however, does not diminish its importance in generating altermagnetic spin splitting~\cite{Autieri2026jpcl}. Remarkably, in hexagonal $\alpha$-MnTe, theoretical studies have shown that SOC generates an additional N\'eel-vector-dependent, d-wave-like spin splitting pattern near the $\Gamma$ in the $\Gamma$-K-M plane, where the nonrelativistic g-wave splitting vanishes~\cite{Yinprl2019,Takahashi2025npj}. The lower symmetry of this texture reflects the coupling between magnetic orientation and the crystal, rather than the sixfold lattice symmetry alone. Recent ARPES measurements from the $(1\bar{1}00)$ surface have resolved this SOC-related d-wave spin splitting and revealed band-dispersion variations associated with different N\'eel-vector orientations~\cite{Usanov2026arxiv}.

The sensitivity of the electronic structure to magnetic orientation~\cite{FariaJunior2023prb,Takahashi2025npj} has direct consequences for magnetotransport, a fundamental property of magnetic materials and a physical basis for spintronic devices. 
Anisotropic magnetoresistance (AMR), for example, can reflect Fermi-surface geometry, electron scattering by impurities, phonons, and magnons, crystal symmetry, and the quantum geometry of electronic bands~\cite{Kriegner2017prb,Betancourt2024npj,Chen2026prb,Zhang2025prlAMR}. 
In altermagnets, distinctive spin-split Fermi surfaces provide a platform for exploring spin transport, while the N\'eel-vector orientation offers a control knob for tuning the response~\cite{Belashchenko2025prl,GonzalezHernandez2021SpinSplitter,Ma2021nc,Lai2025prl,Hirakida2026prb}. 
Analogous to the sensitivity of AMR to magnetization orientation, the intrinsic connection between spin transport and the N\'eel vector provides a means of characterizing and controlling altermagnetic states. 
These advances raise two related questions: How do SOC and staggered exchange jointly generate the additional altermagnetic spin splitting, and how does the N\'eel-vector orientation govern the resulting spin transport?

\begin{figure}[t]
		\centering
		\includegraphics[width=1.0\linewidth]{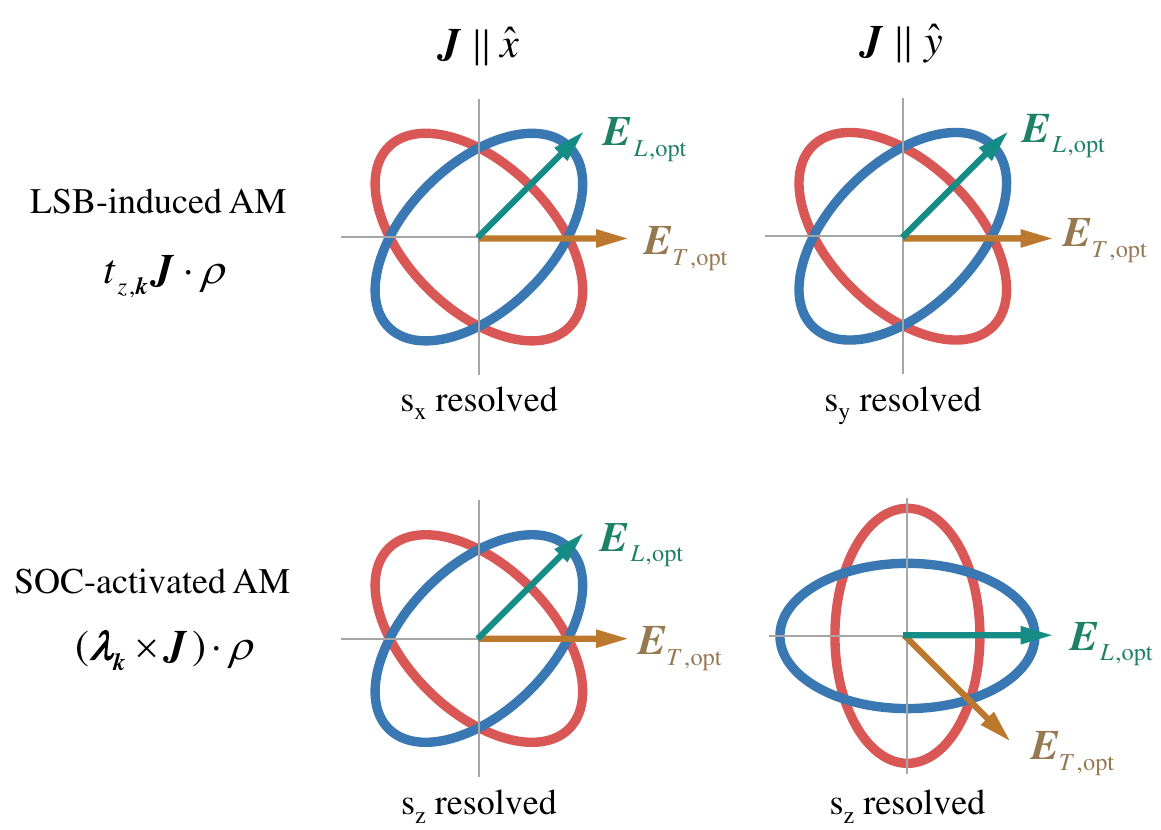}
		\caption{
			Conventional and SOC-activated $d$-wave altermagnetic spin splitting, with schematic terms from Eq.~(\ref{eq:H-eff-alpha}).
			Columns: N\'eel vector along the $x$ (left) and $y$ (right). 
			Top row: local symmetry breaking (LSB), $t_{z,\bm{k}}\propto k_xk_y$, yields a fixed pattern with spin polarization parallel to the N\'eel vector.
			Bottom row: SOC, $\bm{\lambda}_{\bm{k}}\propto(k_x^2-k_y^2,-2k_xk_y,0)$, yields out-of-plane polarization and $45^\circ$ pattern counter-rotation for a $90^\circ$ N\'eel-vector rotation.
			Red/blue denote opposite spin polarizations.
			Green/orange arrows mark electric-field directions maximizing longitudinal/transverse spin conductivities.
		}
		\label{fig:spin_texture_rotation}
	\end{figure}

In this Letter, starting from a symmetry-constrained two-sublattice model~\cite{Roig2024prb}, we derive a low-energy effective model and identify a SOC-activated altermagnetic contribution that can remain finite on momentum planes where the conventional nonrelativistic contribution vanishes. 
For $\alpha$-MnTe, first-principles spin-transport calculations reveal approximate half-angle locking between the N\'eel vector ($\theta_N$) and the electric-field orientations maximizing the signed longitudinal ($\phi_L$) and transverse ($\phi_T$) spin conductivities: $\phi_L\approx45^{\circ}-\theta_N/2$ and $\phi_T\approx-\theta_N/2$.
An analytical effective model traces this locking to a distinctive angular coupling between the SOC-activated spin-splitting pattern and the N\'eel vector.
Specifically, this SOC-activated contribution has two microscopic features, illustrated in Fig.~\ref{fig:spin_texture_rotation}. 
First, its associated spin polarization is perpendicular to the N\'eel vector. Consequently, an in-plane N\'eel vector therefore induces an out-of-plane ($s_z$) spin polarization. 
Second, the angular evolution of the spin-splitting pattern is coupled to the N\'eel-vector orientation, with the rotation relation determined by the momentum dependence of the SOC field. 
In $\alpha$-MnTe, this coupling enters through the phase $2\theta_{\bm{k}}+\theta_N$, where $\theta_{\bm{k}}$ denotes the in-plane orientations of momentum.
The resulting spin-splitting pattern to rotate oppositely to the N\'eel vector by half its rotation angle. 

Our findings elucidates the microscopic mechanism of SOC-activated altermagnetic spin splitting and provides an explanation for the experimentally observed SOC-related splitting. By further establishing half-angle locking between the optimal spin-transport orientations and the N\'eel vector, it provides a theoretical basis for probing N\'eel-vector orientation through angle-dependent spin transport and suggests a route to controlling spin currents in altermagnets.
	
	\textit{SOC-activated altermagnetic spin splitting.---}
	To establish a unified description of the N\'eel-vector-dependent spin splitting, we start from the symmetry-constrained two-sublattice model introduced by~\cite{Roig2024prb},
	\begin{equation}
		H=\varepsilon_{0,\bm{k}}+t_{x,\bm{k}}\tau_x+t_{z,\bm{k}}\tau_z+\tau_y\bm{\lambda}_{\bm{k}}\cdot\bm{\sigma}+\tau_z\bm{J}\cdot\bm{\sigma}
	\end{equation}
	where $\tau_i$ and $\sigma_i$ act in sublattice and physical-spin spaces, respectively. $\varepsilon_{0,\bm{k}}$ represents a sublattice-independent dispersion, $t_{x,\bm{k}}$ is the intersublattice hopping coefficient, $t_{z,\bm{k}}$ is the intrasublattice hopping coefficient, $\bm{\lambda}_{\bm{k}}$ describes the symmetry-allowed SOC, and $\bm{J}=J\bm{n}$ represents the N\'eel order parameter. Here, we resolve the joint SOC–exchange contribution to the effective spin texture and its continuous dependence on the N\'eel-vector orientation.
	
	We treat the exchange field nonperturbatively by diagonalizing $H_0=t_{x,\bm{k}}\tau_x+J\tau_z\bm{n}\cdot\bm{\sigma}$.
	Its two doubly degenerate branches are separated by $2R_{\bm{k}}$, where $R_{\bm{k}}=\sqrt{t_{x,\bm{k}}^2+J^2}$.
	Projecting onto branch $\alpha=\pm1$ and retaining terms linear in $t_z$ and SOC gives
	\begin{equation}\label{eq:H-eff-alpha}
		H_{\text{eff},\alpha}=\epsilon_{\alpha}(\bm{k})\rho_0+\left[\frac{\alpha t_{z,\bm{k}}}{R_{\bm{k}}}\bm{J}-\frac{\bm{\lambda}_{\bm{k}}\times\bm{J}}{R_{\bm{k}}}\right]\cdot\bm{\rho},
	\end{equation}
	where $\epsilon_{\alpha}(\bm{k})=\varepsilon_{0,\bm{k}}+\alpha R_{\bm{k}}$.
	The expansion is valid for $|t_{z,\bm{k}}|+|\bm{\lambda}_{\bm{k}}|\ll R_{\bm{k}}$, with corrections beyond first order neglected. The matrices $\rho_i$ act within the selected doublet. The physical-spin operators are transformed consistently with the Hamiltonian, as detailed in the Supplemental Materials. Eq.~(\ref{eq:H-eff-alpha}) separates the contribution proportional to $t_z\bm{J}$ from a joint SOC-exchange contribution that can remain finite where $t_z=0$.
	
	\begin{figure}[t]
		\centering
		\includegraphics[width=1.0\linewidth]{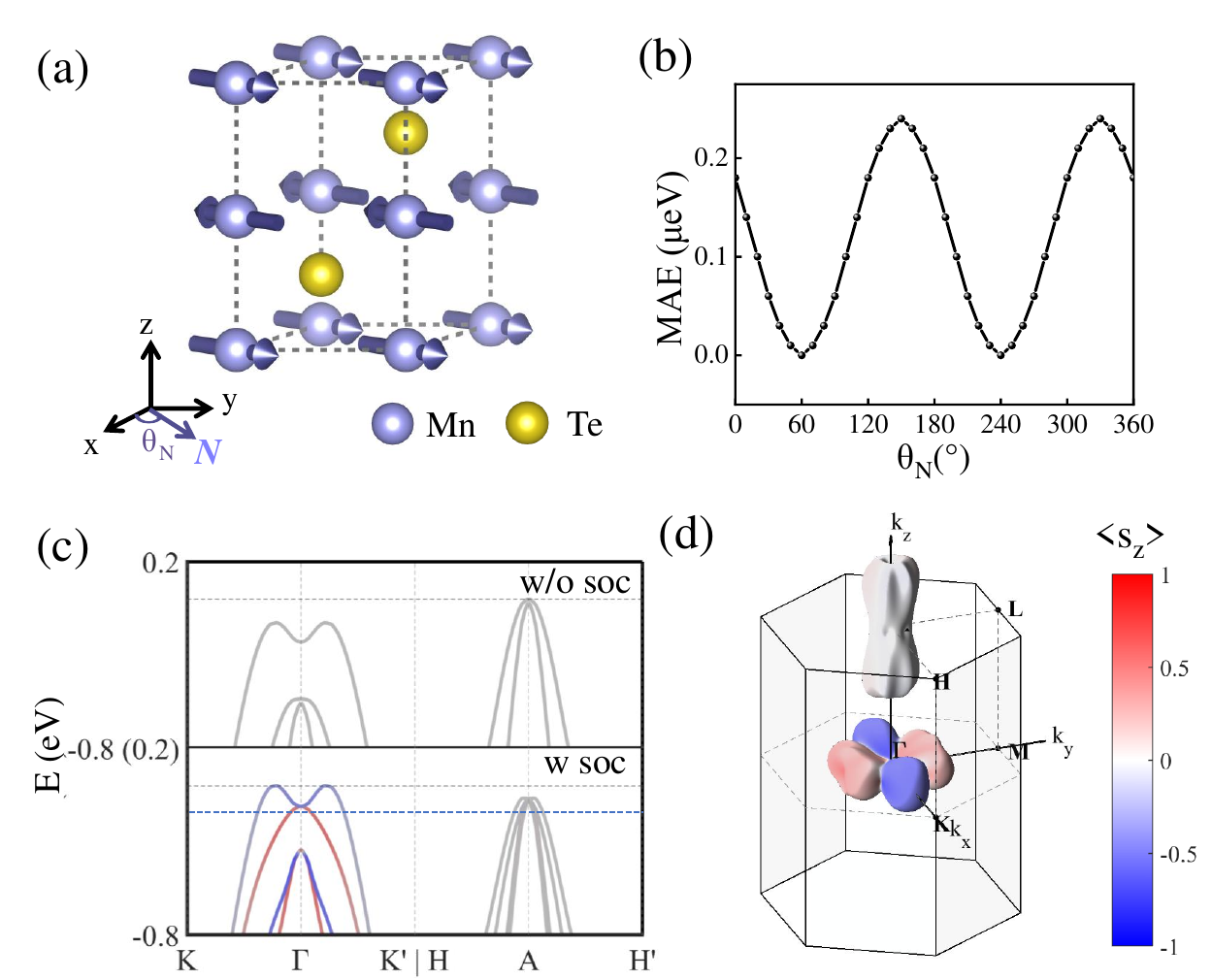}
		\caption{
			(a) Crystal structure and in-plane antiferromagnetic order of hexagonal $\alpha$-MnTe; $\mathbf{N}$ and $\theta_N$ denote the N\'eel vector and its azimuth relative to $x$.
			(b) In-plane magnetic anisotropy energy (MAE) versus $\theta_N$.
			(c) Band structures without SOC (upper) and with SOC (lower) along $K$--$\Gamma$--$K'$ and $H$--$A$--$H'$, with energies referenced to the valence-band maximum.
			The SOC-split bands near $\Gamma$ are colored by the out-of-plane spin polarization $\langle s_z\rangle$.
			(d) Constant-energy surface at $E=-0.14$ eV, marked by the blue dashed line in (c), colored by $\langle s_z\rangle$.
			In (c) and (d), $\theta_N=90^{\circ}$.}
		\label{fig:fig1}
	\end{figure}
	
	\begin{figure*}
		\centering
		\includegraphics[width=0.95\linewidth]{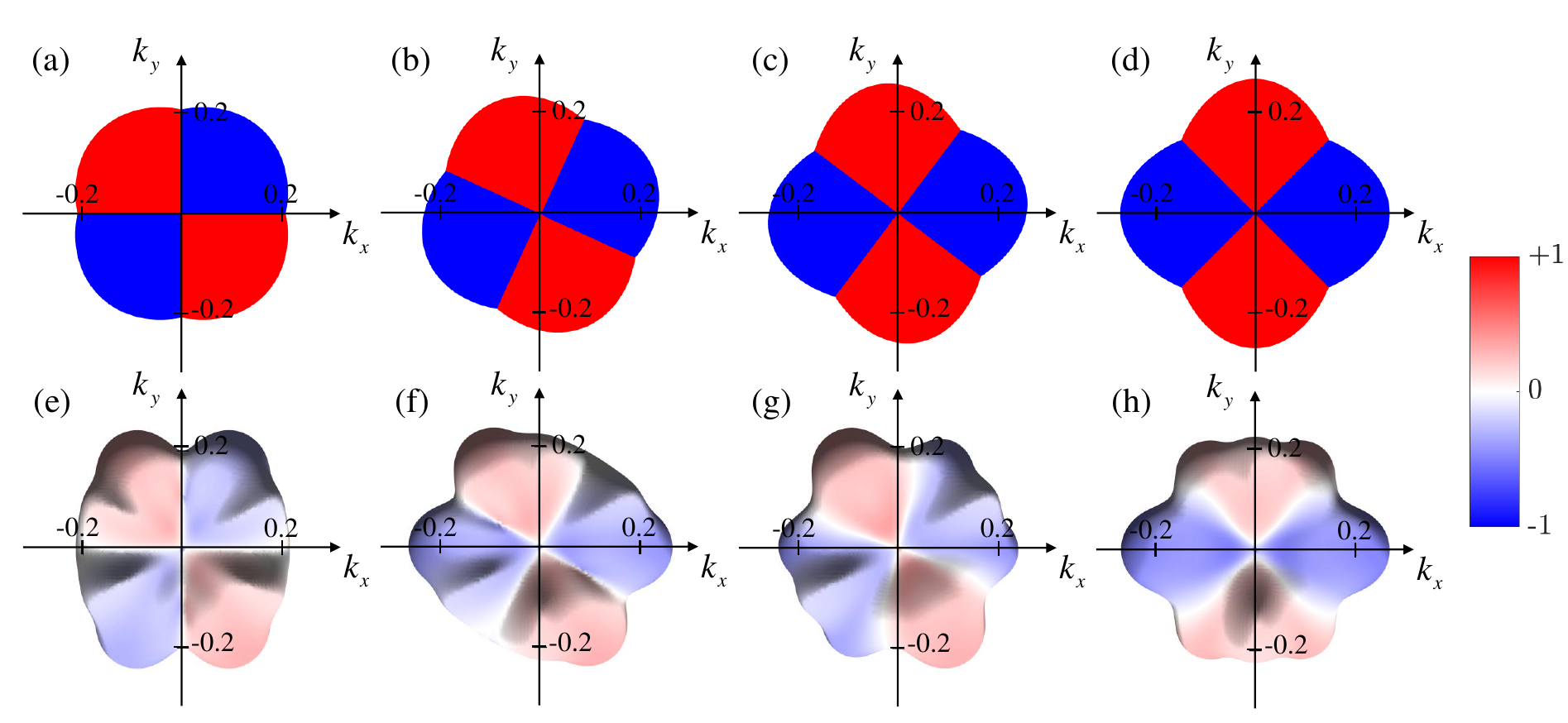}
		\caption{Fermi surfaces projected onto the ($k_x,k_y$) plane, obtained from (a--d) the effective model and (e--h) first-principles calculations for N\'eel-vector angles $\theta_N=0^{\circ}$ (a,e), $30^{\circ}$ (b,f), $60^{\circ}$ (c,g), and $90^{\circ}$ (d,h). Colors denote the effective doublet polarization $\langle \rho_z\rangle$ in (a--d) and the out-of-plane spin polarization $\langle s_z\rangle$ in (e--h). The effective-model parameters are $t=1$, $t_J=0.2$ and $E_F=0.05$. The first-principles results use energy $0.18$ eV below the VBM.}
		\label{fig:isosurface}
	\end{figure*}
	
	For the space-group-194 model relevant to $\alpha$-MnTe, $t_z$ vanishes in the $k_z=0$ plane, whereas the leading in-plane SOC near $\Gamma$ is 
	\begin{equation}\label{eq:lambda}
		\bm{\lambda_k}=\frac{9\lambda_1}{4}(k_x^2-k_y^2,-2k_xk_y,0)+O(k_{\parallel}^4).
	\end{equation}
	For an in-plane N\'eel vector $\bm{J}=J(\cos\theta_N,\sin\theta_N,0)$, Eq.~(\ref{eq:H-eff-alpha}) therefore reduces to
	\begin{equation}\label{eq:H-eff-Gamma}
		H_{\Gamma,\text{eff},\alpha}=\epsilon_{\alpha}(\bm{k})\rho_0+t_Jk_{\parallel}^2\sin(2\theta_{\bm{k}}+\theta_N)\rho_z.
	\end{equation}
	where $t_J=-\frac{9\lambda_1J}{4R_{\Gamma}}$. This expression unifies the $k_xk_y$ and $k_x^2-k_y^2$ structures obtained for the two high-symmetry N\'eel-vector orientations studied by Takahashi et al~\cite{Takahashi2025npj}. It further establishes their continuous angular connection. To leading order, the physical out-of-plane spin is proportional to $\rho_z$, with a projection factor $t_x/R_{\bm{k}}$. Consequently, rotating the N\'eel vector shifts the nodal directions of the spin texture according to $\Delta\theta_{\text{node}}=-\Delta\theta_N/2$. This SOC-exchange-induced angular structure provides the microscopic basis for the transport-axis locking examined below.

	\textit{Electronic structure.---}
	Hexagonal $\alpha$-MnTe ($P6_3/mmc$) hosts in-plane antiparallel moments on its two Mn sublattices [Fig.~\ref{fig:fig1}(a)].
	First-principles calculations used density functional theory (DFT) with the projector augmented-wave (PAW) method in VASP~\cite{Kresse1996,Kresse1999} and the PBE functional~\cite{Perdew1996}.
	Mn-$3d$ correlations were treated within DFT+$U$, with $U=4.0$ eV and Hund exchange $J_h=0.9$ eV~\cite{Dudarev1998,Takahashi2025npj}.
	We used lattice constants of $a=4.171$~\AA\ and $c=6.686$~\AA.
	The weak in-plane magnetic anisotropy energy (MAE, $<1$ meV) [Fig.~\ref{fig:fig1}(b)] suggests that fields, spin torques, or strain can reorient the N\'eel vector.
	
	For spin-texture and transport calculations, we constructed a Wannier Hamiltonian for the Te-$p$-dominated states near the valence-band maximum (VBM), using 12 spinor Wannier functions in Wannier90~\cite{Mostofi2008,Mostofi2014,Marzari2012}.
	Comparison with the SOC-free bands reveals additional spin splitting near $\Gamma$ with opposite out-of-plane spin polarizations [Fig.~\ref{fig:fig1}(c)].
	At $E=-0.14$ eV relative to the VBM, the $\Gamma$ pocket exhibits a $d$-wave-like out-of-plane spin texture, whereas the $A$ pocket is nearly unpolarized along $z$ [Fig.~\ref{fig:fig1}(d)].
	This identifies the $\Gamma$ states as a channel for the SOC--exchange-induced spin texture.
	
	
	For comparison with first-principles results and analytical transport calculations, we approximate the spin-independent dispersion in Eq.~(\ref{eq:H-eff-Gamma}) as $\epsilon_{\alpha}(\bm{k})\simeq tk^2$, where $k^2=k_x^2+k_y^2+k_z^2$ and $t$ is the isotropic dispersion coefficient. The isotropic form is adopted for analytical tractability rather than required by hexagonal symmetry. 
	Figure~\ref{fig:isosurface} compares the model [panels (a)--(d)] and first-principles [panels (e)--(h)] textures for $\theta_N=0^{\circ}$, $30^{\circ}$, $60^{\circ}$, and $90^{\circ}$.
	The phase $2\theta_{\bm{k}}+\theta_N$ couples momentum and magnetic orientation, making the nodal directions and polarization pattern counter-rotate by half the N\'eel-vector rotation angle.
	The first-principles textures reproduce this leading angular evolution, with additional Fermi-surface deformations and polarization variations beyond the isotropic quadratic model.
	The agreement therefore concerns the angular structure, rather than a quantitative reproduction of the full Fermi-surface geometry or polarization amplitude.
	We therefore use the minimal model to isolate this angular coupling and the Wannier Hamiltonian for material-specific longitudinal and transverse spin transport.
	
	

	\textit{Formalism of spin transport.---}
	To quantify the transport consequences of the N\'eel-vector-dependent spin texture, we calculate the in-plane spin-conductivity tensor for out-of-plane spin polarization. We define the spin-velocity operator as $\hat{v}_i^z=\{\hat{v}_i,\hat{\sigma}_z\}/2$, and use the charge-normalized spin current, $J_i^z=-e\sum_n\int[d\bm{k}]f_{n\bm{k}}v_{n,i}^z$, where $e>0$ and $[d\bm{k}]=d^3\bm{k}/(2\pi)^3$. Within the constant-relaxation-time approximation, the intraband contribution is 
	\begin{equation}\label{eq:sigma-intra}
		\sigma^{s_z,\text{intra}}_{ij}=e^2\tau \sum_n\int[d\bm{k}] v^{z}_{n,i}v_{n,j}(\bm{k})\left(-\frac{\partial f_{FD}}{\partial E_{n\bm{k}}}\right),
	\end{equation}
	Here, $f_{FD}$ is the equilibrium Fermi–Dirac distribution, and \(v_{n,j}=\langle u_{n\mathbf k}|\hat v_j|u_{n\mathbf k}\rangle\) and \(v_{n,i}^{z}=\langle u_{n\mathbf k}|\hat v_i^{z}|u_{n\mathbf k}\rangle\) are the band-resolved velocity and spin velocity, respectively. We adopt a constant carrier relaxation time of $\tau=10^{-13}$s, consistent in order of magnitude with mobility-derived estimates for Li-doped MnTe~\cite{Zheng2019paramagnon}. The intrinsic interband contribution is 
	\begin{equation}\label{eq:sigma-inter}
		\sigma^{z,\text{inter}}_{ij}=-\frac{e^2}{\hbar}\sum_{n}\int[d\bm{k}]f_{FD}(E_{n\bm{k}})\Omega_{n,ij}^{z},
	\end{equation}
	where $\Omega_{n,ij}^{z}$ denotes the spin Berry curvature. Both contributions are evaluated using the Wannier Hamiltonian. Derivation and implementation details are provided in the Supplemental Materials.
	
	To identify the preferred transport directions, we decompose the crystal-frame tensor as
	\begin{equation}\label{eq:sigma-unrotated}
		\sigma^{z}=\begin{pmatrix}
			f_0+f_z&f_x+f_y\\ f_x-f_y&f_0-f_z
		\end{pmatrix}
	\end{equation}
	Here, $f_0$ is the isotropic diagonal component, $f_x$ and $f_z$ describe the symmetric anisotropic response, and $f_y$ is the antisymmetric component.
	
	We apply the electric field along an in-plane direction $x'$ at an angle $\phi$ to the crystal $x$ axis, with $y'$ rotated counterclockwise by $90^{\circ}$ from $x'$. The longitudinal and transverse spin conductivities are then
	\begin{equation}\begin{aligned}
			\sigma^{z}_{x'x'}(\phi)=&f_0+f_x\sin2\phi+f_z\cos2\phi,\\
			\sigma^{z}_{y'x'}(\phi)=&f_x\cos2\phi-f_z\sin2\phi-f_y.
	\end{aligned}\end{equation}
	Then, we define $\phi_L$ and $\phi_T$ as the electric-field orientations maximizing the signed longitudinal and transverse responses, respectively. For $A=\sqrt{f_x^2+f_z^2}$, these orientations are $\phi_L={\rm atan2}(f_x,f_z)/2$ and $\phi_T={\rm atan2}(-f_z,f_x)/2$ with maximum response $f_0+A$ and $A-f_y$. When $A$ vanishes, both responses are independent of $\phi$. The background $f_0$ and $f_y$ shift the responses without changing their optimal orientations.

	\begin{figure}[t]
		\centering
		\includegraphics[width=1.0\linewidth]{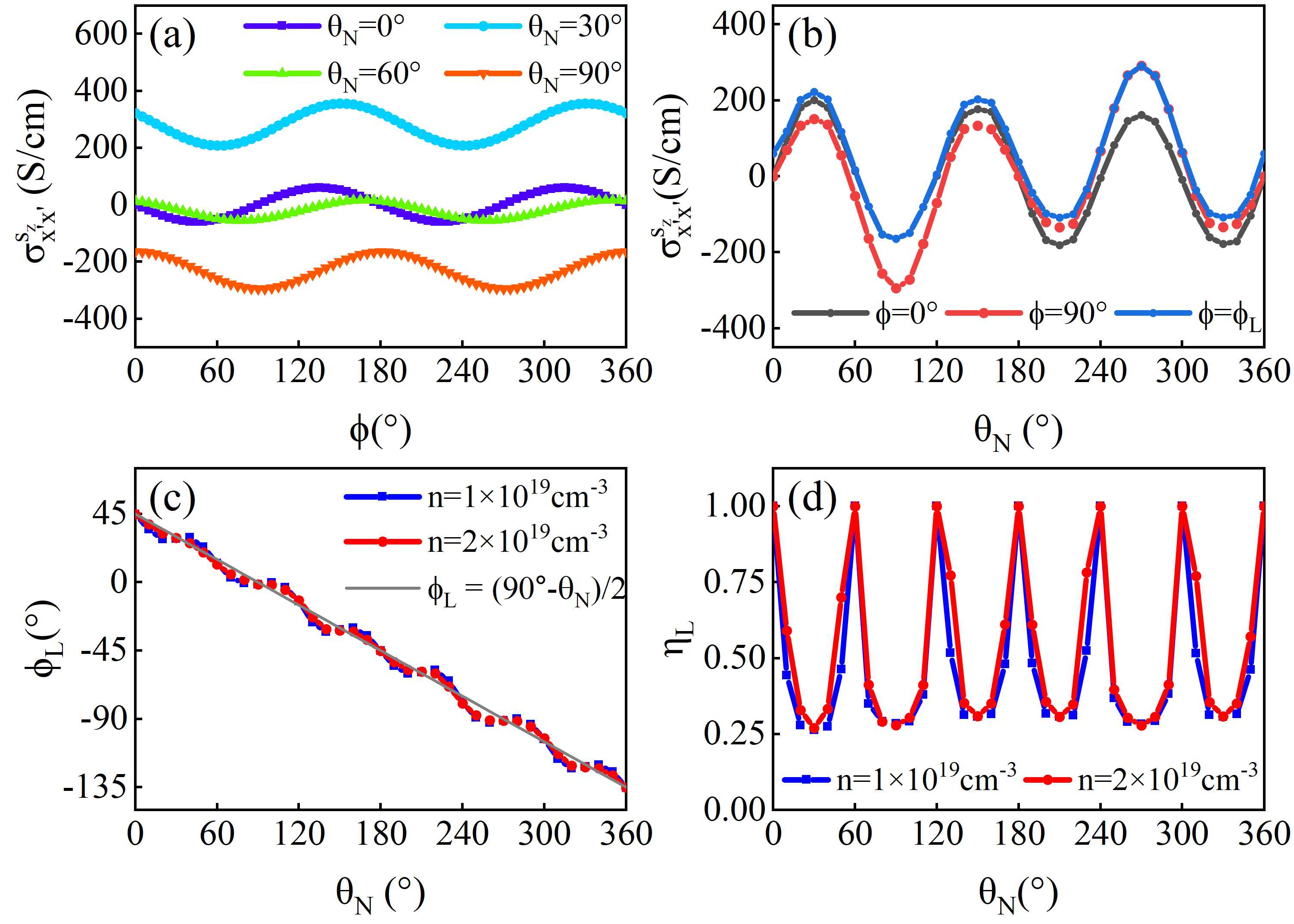}
		\caption{
			(a) Longitudinal spin conductivity $\sigma_{x'x'}^{s_z}$ as a function of $\phi$ for N\'eel vector angles $\theta_N=0^{\circ}$, $30^{\circ}$, $60^{\circ}$, and $90^{\circ}$.
			(b) $\sigma_{x'x'}^{s_z}$ versus $\theta_N$ for $\phi=0^{\circ}$, $90^{\circ}$, and $\phi_L$.
			(c) Optimal orientation angle $\phi_L$ and (d) longitudinal spin transport anisotropy ratio $\eta_L$ as functions of $\theta_N$ for carrier concentrations $n=1\times10^{19}$ and $2\times10^{19}~\mathrm{cm}^{-3}$.
			In (a) and (b), the carrier concentration is fixed at $n=1\times10^{19}~\mathrm{cm}^{-3}$.}
		\label{fig:fig2}
	\end{figure}

	\begin{figure}[t]
		\includegraphics[width=1.0\linewidth]{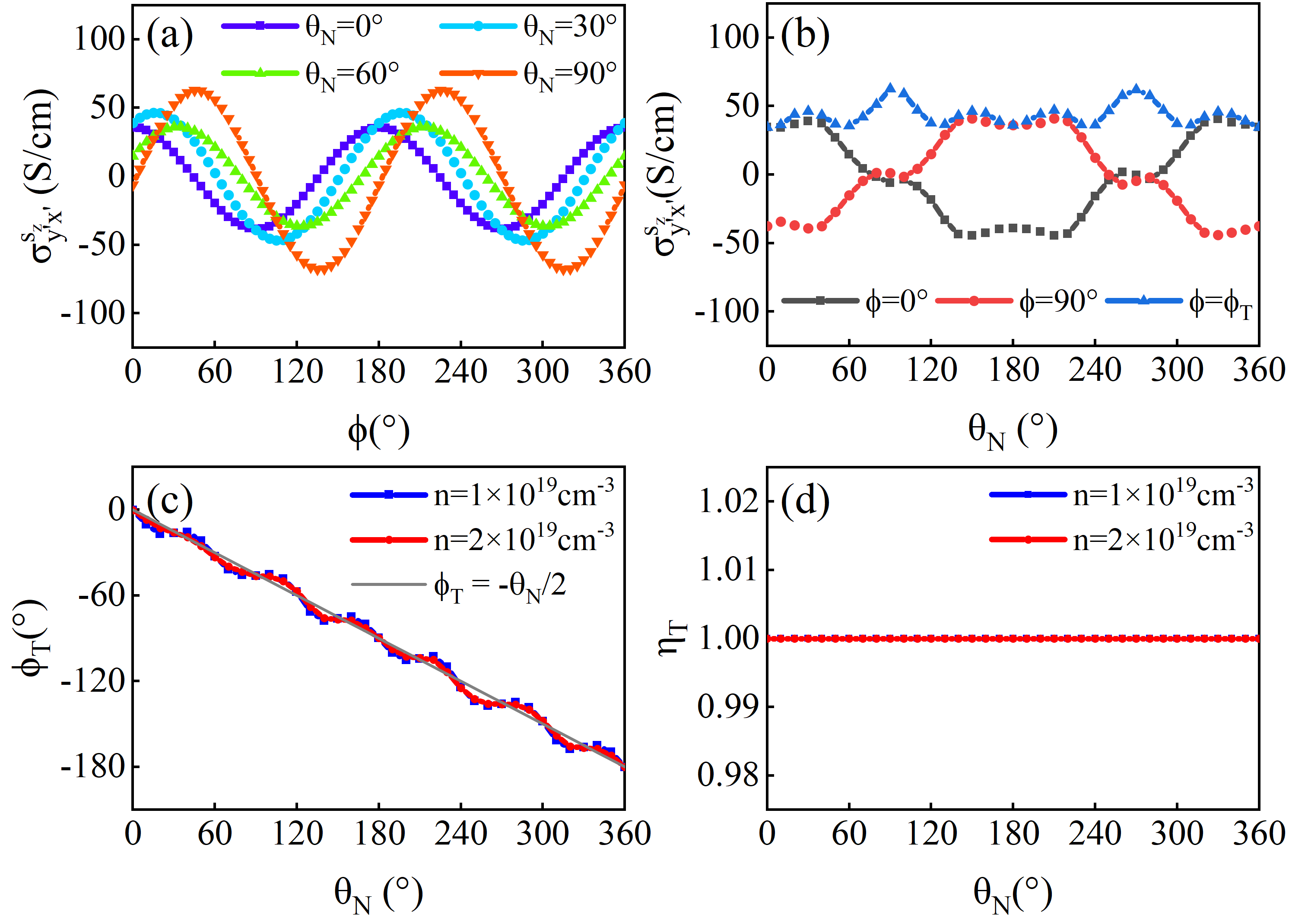}
		\caption{
			(a) Transverse spin conductivity $\sigma_{y'x'}^{s_z}$ as a function of $\phi$ for N\'eel vector angles $\theta_N=0^{\circ}$, $30^{\circ}$, $60^{\circ}$, and $90^{\circ}$.
			(b) $\sigma_{y'x'}^{s_z}$ versus $\theta_N$ for $\phi=0^{\circ}$, $90^{\circ}$, and $\phi_T$.
			(c) Optimal orientation angle $\phi_T$ and (d) transverse spin transport anisotropy ratio $\eta_T$ as functions of $\theta_N$ for carrier concentrations $n=1\times10^{19}$ and $2\times10^{19}~\mathrm{cm}^{-3}$.
			In (a) and (b), the carrier concentration is fixed at $n=1\times10^{19}~\mathrm{cm}^{-3}$.}
		\label{fig:fig3}
	\end{figure}

	\textit{Optimal spin-transport direction.---}
	Figure~\ref{fig:fig2} and \ref{fig:fig3} present the longitudinal and transverse spin conductivities calculated using the Wannier Hamiltonian. We vary both the in-plane N\'eel-vector orientation $\theta_N$ and the electric-field direction $\phi$ to determine how the evolving spin texture controls the preferred transport orientations.
	
	For both channels, the conductivity exhibits a $180^{\circ}$-periodic modulation with $\phi$ [Figs.~\ref{fig:fig2}(a) and \ref{fig:fig3}(a)], consistent with the tensor rotation derived above. The longitudinal curves show appreciable N\'eel-vector-dependent offsets, indicating a varying isotropic component $f_0$. By contrast, the transverse curves have small offsets, consistent with a weak antisymmetric component $f_y$. The two channels therefore share a symmetric anisotropic contribution while differing in their angle-independent backgrounds.
	
	Figures~\ref{fig:fig2}(b) and \ref{fig:fig3}(b) compare the responses along the fixed crystallographic directions $\phi=0^{\circ}$ and $90^{\circ}$ with those obtained at the optimal orientation $\phi_L$ and $\phi_T$. As the N\'eel vector rotates, the response along a fixed direction varies and can change sign. Selecting the optimal orientation instead follows the maximum signed response. Although a fixed crystallographic direction can approach the optimum for particular N\'eel-vector orientations, it does not generally track the maximum throughout the rotation of N\'eel vector. This distinction is particularly evident for the transverse response, highlighting the importance of adjusting the measurement direction when comparing spin-current generation across magnetic orientations.
	
	The central result is the continuous evolution of the optimal orientations shown in Figs.~\ref{fig:fig2}(c) and \ref{fig:fig3}(c). The numerical results approximately follow
	\begin{equation}\label{eq:locking-longitudinal-transverse}
		\phi_L\approx45^{\circ}-\frac{\theta_N}{2},
		\ \ \ \
		\phi_T\approx-\frac{\theta_N}{2}.
	\end{equation}
	Thus, the optimal transport orientations rotate oppositely to the N\'eel vector, with approximately half its rotation angle. This behavior is found at both carrier concentrations examined, $n=1\times10^{19}$ and $2\times10^{19}$ cm$^{-3}$. The $45^{\circ}$ separation between the longitudinal and transverse optima follows from tensor geometry, whereas their approximately linear dependence on $\theta_N$ reflects the evolution of the symmetric anisotropic coefficients $f_x$ and $f_z$. 
	
	The anisotropy ratios in Figs.~\ref{fig:fig2}(d) and \ref{fig:fig3}(d) further distinguish the two channels. Defining $\eta=(\sigma_{\text{max}}-\sigma_{\text{min}})/(|\sigma_{\text{max}}|+|\sigma_{\text{min}}|)$ for each angular response, the longitudinal ratio varies with $\theta_N$ reflecting the changing balance between the isotropic background and symmetric anisotropy. The transverse anisotropy ratio remains close to unity, consistent with the nearly zero-centered angular oscillations in Fig.~\ref{fig:fig3}(a). These channel-dependent backgrounds affect the response amplitudes but do not determine the optimal orientations.

	\textit{Analytical origin of half-angle locking.---}
	To identify the origin of the numerically observed locking between the optimal and N\'eel-vector orientations, we evaluate the spin conductivity of the isotropic low-energy model. Its two eigenvalues are $E_{\pm\bm{k}}=tk^2\pm t_Jk_{\parallel}^2\sin(2\theta_{\bm{k}}+\theta_N)$. The eigenstates have definite $\rho_z$ polarization. Within this minimal model, we define the effective spin-velocity operator as $\hat{v}_i^z=\{\hat{v}_i,\rho_z\}/2$. Both the velocity and effective spin-velocity operators are diagonal in the same basis, so the intrinsic interband contribution vanishes. The remaining intraband response can be evaluated analytically at zero temperature. For $t>|t_J|$ and a Fermi energy $E_F>0$ measured from the band edge, we obtain 
	\begin{equation}
		\sigma_{\text{eff}}^z=C\begin{pmatrix}
			\sin\theta_N&\cos\theta_N\\
			\cos\theta_N&-\sin\theta_N
		\end{pmatrix}
		\ \ \text{with}\ \ 
		C=\frac{2e^2\tau E_F^{3/2}t_J}{3\pi^2\hbar^2\sqrt{t(t^2-t_J^2)}}.
	\end{equation}
	The derivation is provided in the supplementary materials. The prefactor $C$ determines the response amplitude, whereas the matrix determines its angular structure. In particular, the isotropic diagonal and antisymmetric components vanish in this model, while $f_x=C\cos\theta_N$ and $f_z=C\sin\theta_N$.
	
	Rotating the measurement direction by $\phi$ gives
	\begin{equation}
		\sigma^z_{x'x',\text{eff}}=C\sin(2\phi+\theta_N)
		\ \ \text{and}\ \
		\sigma^z_{y'x',\text{eff}}=C\cos(2\phi+\theta_N).
	\end{equation}
	The momentum-space phase $(2\theta_{\bm{k}}+\theta_N)$ thus produces a corresponding measurement-angle dependence through $(2\phi+\theta_N)$, which is consistent with the numerical results in Eq.~(\ref{eq:locking-longitudinal-transverse}). Within this quadratic model, varying $E_F$ or a common relaxation time changes the amplitude without changing the optimal orientations. The analytical result therefore explains why the two carrier concentrations examined in the Wannier calculations exhibit similar angular locking. The finite longitudinal background and deviations from the ideal relation in the material-specific calculations arise beyond the traceless isotropic model. The analytical model isolates the angular coupling responsible for the locking, while the Wannier results establish its approximate realization in $\alpha$-MnTe.

	\textit{Summary and Discussion.---}
	In the conventional nonrelativistic picture, altermagnetic spin splitting arises from the interplay between staggered exchange and sublattice-dependent hopping permitted by reduced local point symmetry. The anisotropic local environments produce different dispersions on the two magnetic sublattices, which combine with opposite exchange fields to lift spin degeneracy at a given momentum. Our results elucidate a relativistic channel in which SOC and staggered exchange jointly generate an altermagnetic spin texture where conventional contribution vanishes. In the $\Gamma-K-M$ plane of $\alpha$-MnTe, their coupling produces an out-of-plane d-wave altermagnetic spin texture whose angular pattern depends continuously on the N\'eel-vector orientation. This channel complements the nonrelativistic g-wave altermagnetism of MnTe and connects its microscopic SOC-exchange coupling to controllable low-energy spin polarization.
	
	The resulting altermagnetic spin splitting evolution has a direct transport consequence. The first-principles and Wannier calculations reveal approximate half-angle locking of the optimal longitudinal and transverse spin-transport directions. The analytical quadratic model traces this behavior to the coupled momentum and N\'eel-vector phase, which determines the orientation of the symmetric anisotropic conductivity. Within this model, the carrier density and a common relaxation time change the response amplitude without changing the optimal orientations. The material-specific calculations additionally contain band-structure corrections and angle-independent backgrounds, supporting an approximate realization of the ideal locking relation. Together, these results identify the N\'eel vector as a control parameter for directional altermagnetic spin-current generation.
	
	This continuous description connects the SOC-exchange dependence of the minimal model of Roig et al.~\cite{Roig2024prb} with the high-symmetry effective Hamiltonian obtained by Takahashi et al.~\cite{Takahashi2025npj}, and extends them to the evolution of spin texture and optimal transport directions. Recent ARPES observation of SOC-related d-wave splitting and N\'eel-vector-dependent band dispersion provide an experimental setting for testing these predictions~\cite{Usanov2026arxiv}. Measurements with controlled magnetic orientation could examine the predicted counter-rotation of the spin texture and the accompanying transport-axis locking, thereby testing the connection between microscopic spin polarization and directional spin-current generation.
	
	Beyond MnTe, symmetry-allowed SOC fields provides a route to identifying analogous N\'eel-vector-dependent altermagnetic textures. Their realization requires examining the physical-spin projection and competing low-energy terms for each space group, Wyckoff position, and magnetic orientation. Analogous SOC-exchange couplings need not produce the same half-angle relation. Within these boundaries, the present framework links microscopic symmetry and exchange to tunable altermagnetic spin textures, providing a basis for exploring orientation-dependent spin transport and transport-based probes of magnetic order.

	\textit{Acknowledgments.---}	
	This work was supported by the National Natural Science Foundation of China (Grants No.12595331), the National Key Research and Development Program of the Ministry of Science and Technology of China (No.2022YFA1602601),the National Natural Science Foundation of China (No.12574048, 12574257 and 12304214). The DFT calculations in this work are supported by the Beijing Super Cloud Computing Center (BSCC).

	\bibliography{references}

\end{document}